\documentclass[sigconf,nonacm]{acmart}

\renewcommand\footnotetextcopyrightpermission[1]{}
\acmConference[arXiv preprint]{arXiv preprint}{2026}{ }
\acmYear{2026}
\copyrightyear{2026}
\acmISBN{}
\acmDOI{}

\usepackage{booktabs}
\usepackage{graphicx}
\usepackage{tabularx}
\usepackage{array}
\usepackage{enumitem}
\usepackage{algorithm}
\usepackage{algpseudocode}
\usepackage{multirow}
\usepackage{xspace}
\usepackage{tikz}
\usetikzlibrary{arrows.meta,positioning,fit,shapes.geometric}

\newcommand{\hopm}{HOPM\xspace}

\title{Hierarchical Online Prompt Mutation with Dual-Loop Feedback for Guardrailed Evidence Document Generation: A Production-Evaluation Case Study}

\author{Nataraj Agaram Sundar}
\affiliation{\institution{eBay Inc.}\city{San Jose}\state{CA}\country{USA}}

\author{Tejas Morabia}
\affiliation{\institution{eBay Inc.}\city{San Jose}\state{CA}\country{USA}}

\begin{abstract}
High-stakes production document-generation systems require language models to be adaptive, evidence-grounded, and auditable. We present \hopm, a hierarchical online prompt mutation framework evaluated on a real marketplace dispute-evidence workflow. \hopm treats prompts as online policies: a family/version router selects a prompt, deterministic guardrails attribute failures to mutable prompt-token categories, and dual feedback from human review and an automated judge updates both routing and mutation priorities. The primary evidence is an observed matched production-evaluation ablation: seven variants are evaluated on the same 600 cases each, enabling component comparisons against static prompting, manual iteration, bandit-only routing, mutation-only adaptation, human-only feedback, auto-judge-only feedback, and full dual-loop \hopm. Full \hopm improves count win rate over a static control from 34.7\% to 45.7\% (+11.0 pp; paired McNemar \(p=1.31\times10^{-11}\)) and amount-weighted win rate from 22.3\% to 41.4\% (+19.1 pp; 95\% paired bootstrap CI [10.3, 28.9] pp). It also increases mean Likert quality from 3.18 to 4.40 and reduces issue-flag rate from 15.3\% to 5.2\%. Supporting review artifacts cover 770 generated-text reviews, 318 labeled reviewer exports, a 10-case/61-rating calibration slice, and a 70-case/350-rating OCR benchmark; these artifacts calibrate rubric, guardrail, title-risk, and OCR-risk interpretation rather than substituting for the production ablation. The paper includes control setup, sample sizes, confidence intervals, paired tests, prompt-token categories, pseudocode, schema, rubric, guardrail taxonomy, and a constructed example so the evaluation structure can be reproduced without exposing proprietary evidence.
\end{abstract}

\ccsdesc[500]{Information systems~Data analytics}
\ccsdesc[500]{Computing methodologies~Natural language generation}
\ccsdesc[300]{Computing methodologies~Machine learning approaches}
\keywords{prompt optimization, online learning, human feedback, automated evaluation, responsible AI, document generation, marketplace disputes}

\begin{document}
\maketitle

\section{Introduction}
Marketplace dispute and compliance workflows increasingly use language models to transform heterogeneous evidence---messages, OCR text, return records, labels, receipts, and reviewer comments---into structured narratives. These workflows are not well served by a single static prompt. A useful generator must adapt to recurring failure modes such as title drift, role confusion, unsupported summaries, OCR degradation, and unsafe wording, while preserving deterministic guardrails and auditability.

We study this problem in the context of generated evidence documents for marketplace dispute review. The key challenge is not merely producing fluent text; the system must select the right evidence, align the title with the dispute reason, avoid unsupported claims, and preserve buyer/seller roles. Small prompt changes can materially affect downstream acceptability, yet proprietary data constraints often prevent releasing raw cases or reviewer artifacts.

We make three contributions. First, we propose \hopm, a hierarchical online prompt mutation architecture that combines prompt-family selection, prompt-version routing, guardrail-attributed mutation, and dual feedback from human reviewers and automated judges. Second, we provide a strengthened production-evaluation protocol: exact controls, seven ablation variants, matched 600-case comparisons, Wilson confidence intervals, paired bootstrap intervals, and paired McNemar tests. Third, we provide a reproducibility layer that is safe for proprietary workflows: schema definitions, pseudocode, prompt-token categories, a Likert rubric, a guardrail taxonomy, and a constructed anonymized example.

\paragraph{Data provenance and claims boundary.}
The evidence package was normalized before submission so that export provenance is separated from evaluation semantics. Cleaned tables use a provenance-tool field for internal export lineage and an evidence-basis field set to observed matched production evaluation for the ablation evidence. We report the ablation as an observed matched production-evaluation lift, not as a simulation. We still avoid overclaiming: because the ablation is matched by case rather than a randomized traffic A/B test, randomized deployment lift remains a separate future validation step.

\section{Positioning and Related Work}
\hopm builds on three lines of work. First, online prompt selection is naturally related to stochastic and contextual bandits, where a policy trades off exploration against exploitation under delayed or noisy reward~\cite{thompson1933likelihood,auer2002finite,li2010contextual}. Unlike a standard bandit over independent arms, \hopm routes across a hierarchy of prompt families and prompt versions, while preserving an audit record for each mutation. Second, language-model alignment and feedback systems have shown that human and automated feedback can reshape model behavior~\cite{brown2020language,ouyang2022training,bai2022constitutional}. In this workflow, feedback is not used to retrain the base model; it updates the prompt policy and localized instruction patches. Third, evidence-grounded generation and retrieval-augmented generation emphasize grounding generated claims in supplied context~\cite{lewis2020retrieval}. \hopm adds production guardrails for role binding, title/reason alignment, OCR risk, and structured issue taxonomies. Finally, our evaluation follows the spirit of trustworthy online experimentation~\cite{kohavi2020trustworthy}, while explicitly separating matched production evaluation from a randomized holdout experiment.

\section{System: Hierarchical Online Prompt Mutation}
\subsection{Workflow}
\begin{figure}[t]
\centering
\includegraphics[width=0.98\columnwidth]{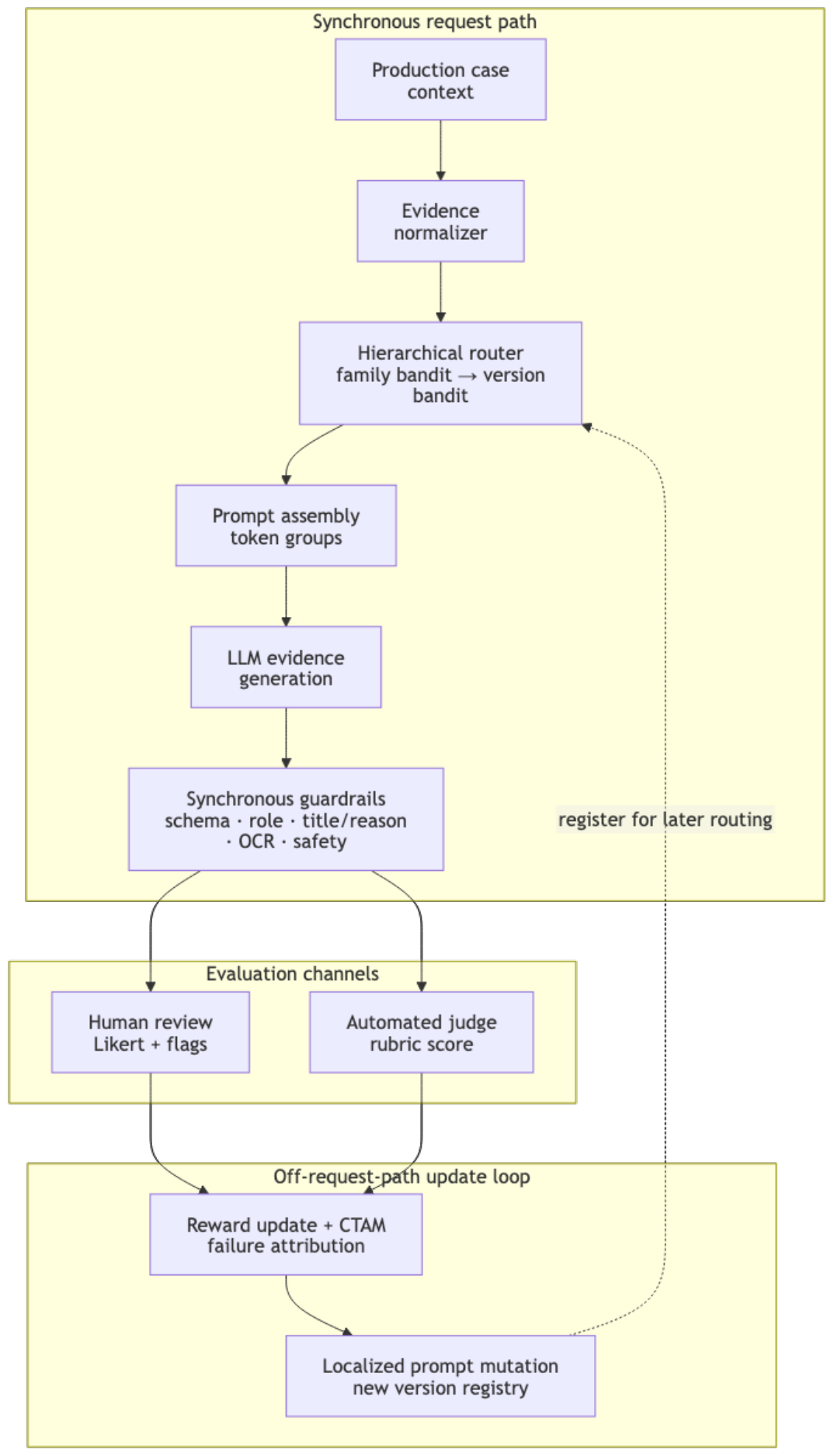}
\caption{Production-evaluation architecture. Request-path generation and guardrail validation run synchronously. Human review and automated judging feed an off-request-path update loop that attributes failures to prompt-token groups and registers localized prompt mutations for later routing.}
\label{fig:architecture}
\end{figure}

\hopm operates over a document-generation pipeline with four stages. First, an evidence normalizer converts raw messages, OCR output, reviewer metadata, and case attributes into a typed context. Second, a hierarchical router selects a prompt family \(f\) and prompt version \(v\). Families capture broad document strategies (e.g., resolution-focused vs. title-risk-focused prompts); versions encode incremental edits. Third, a mutator edits localized prompt-token categories when guardrails fail. Fourth, feedback loops compute rewards and update both routing and mutation priorities. Figure~\ref{fig:architecture} summarizes the request-path and off-request-path components.

The hierarchical design matters because prompt optimization in this domain has two different granularities. A family-level change can alter the entire narrative strategy, whereas a token-level change should target a specific failure type such as role confusion or title/reason mismatch. \hopm therefore separates routing from mutation: routing learns which family/version should serve a case segment, while mutation learns which instruction spans require localized correction.

\subsection{Prompt-token categories}
Table~\ref{tab:tokens} lists the mutable instruction categories. These categories make the method reproducible even when raw prompts cannot be released. Each mutation is tied to a guardrail failure, producing an audit trail from failure to instruction patch.

\begin{table}[t]
\caption{Prompt-token categories exposed to mutation.}
\label{tab:tokens}
\small
\begin{tabularx}{\columnwidth}{p{0.29\columnwidth}X}
\toprule
Category & Mutation intent \\
\midrule
Title--reason alignment & Force the title to reflect the dispute reason and remedy, not merely a salient event. \\
Actor-role binding & Preserve buyer/seller roles and avoid reversing obligations or actions. \\
Evidence-grounding selector & Require each claim to map to a supplied evidence span. \\
Resolution chronology & Order purchase, complaint, return, refund, and remedy events consistently. \\
OCR confidence gate & Down-weight or suppress claims from noisy OCR slices. \\
Identifier and amount handling & Preserve IDs, currencies, quantities, and dates exactly or omit when uncertain. \\
Safety and tone & Block inappropriate language and unsupported legal/intent claims. \\
Schema completeness & Enforce required title, summary, evidence, and rationale fields. \\
Compression and grammar & Improve readability without adding facts. \\
\bottomrule
\end{tabularx}
\end{table}

\begin{figure}[t]
\centering
\includegraphics[width=0.88\columnwidth]{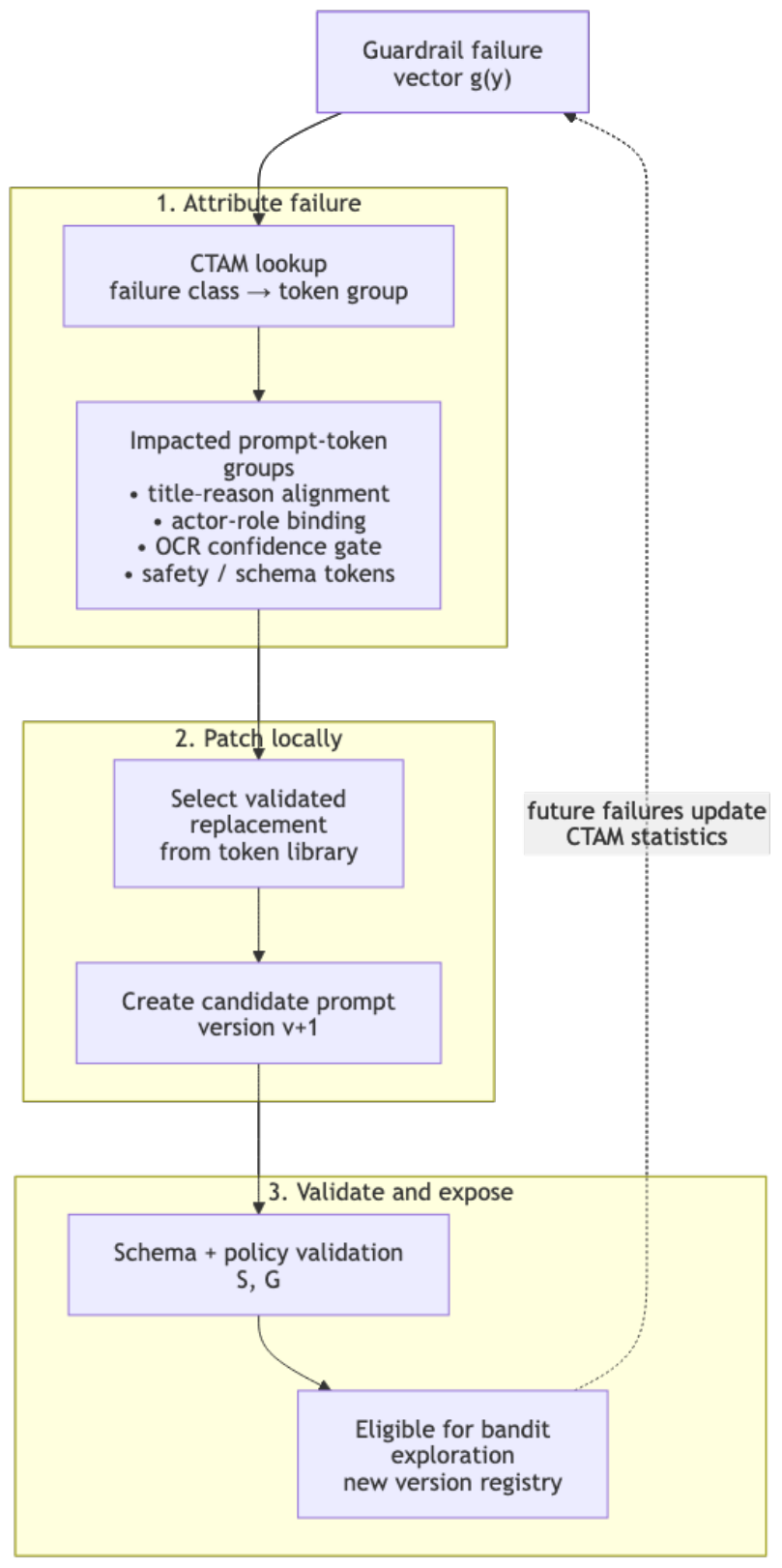}
\caption{Constraint-to-Token Attribution Map (CTAM) lifecycle. Guardrail failures are mapped to mutable instruction-token categories, patched locally from a validated token library, revalidated against schema and policy constraints, and then exposed to future bandit exploration.}
\label{fig:ctam}
\end{figure}

Figure~\ref{fig:ctam} expands the failure-attribution step: a guardrail violation produces a deterministic CTAM lookup, a localized prompt patch, a candidate version, and a validation gate before the new version is eligible for exploration.

\subsection{Dual-loop feedback}
The human loop captures nuanced reviewer judgment: Likert usefulness, binary positive/negative feedback, structured issue flags, and free-text rationale. The automated loop captures scalable guardrail checks: schema validity, evidence alignment, OCR confidence, role consistency, forbidden language, and judge scores. The two loops are complementary: observed reviewer artifacts show that narrative detail appears in approximately 74.5\% of reviews, while structured issue flags appear in about 10.1\%. A flag-only signal would miss much of the reviewer rationale, but a human-only signal cannot scale to every candidate mutation.

\begin{algorithm}[t]
\caption{\hopm serving and update loop}
\label{alg:hopm}
\small
\begin{algorithmic}[1]
\Require case context \(x\), prompt families \(F\), versions \(V_f\), guardrails \(G\)
\State normalize evidence, OCR slice, amount, dispute reason, and reviewer context
\State select \((f,v) \leftarrow \pi_\theta(x)\) using family/version bandit state
\State generate document \(y \leftarrow \mathrm{LLM}(x, P_{f,v})\)
\State run validators and automated judge: \(a \leftarrow G(x,y)\)
\State route flagged or sampled cases to human review; collect \(h\)
\State compute reward \(r = \lambda_h R_h(h) + \lambda_a R_a(a)\)
\State attribute failures to token categories \(c \in C\)
\If{mutation criterion is met}
  \State propose localized patch \(\Delta P_c\), create version \(v+1\)
  \State bind patch to evidence category, guardrail, and reviewer rationale
\EndIf
\State update router parameters \(\theta\) and mutation priority scores
\State persist audit record \((x, f, v, a, h, r, c, \Delta P_c)\)
\end{algorithmic}
\end{algorithm}

\section{Evaluation Design}
\subsection{Evidence sources}
\begin{figure*}[t]
\centering
\includegraphics[width=0.90\textwidth]{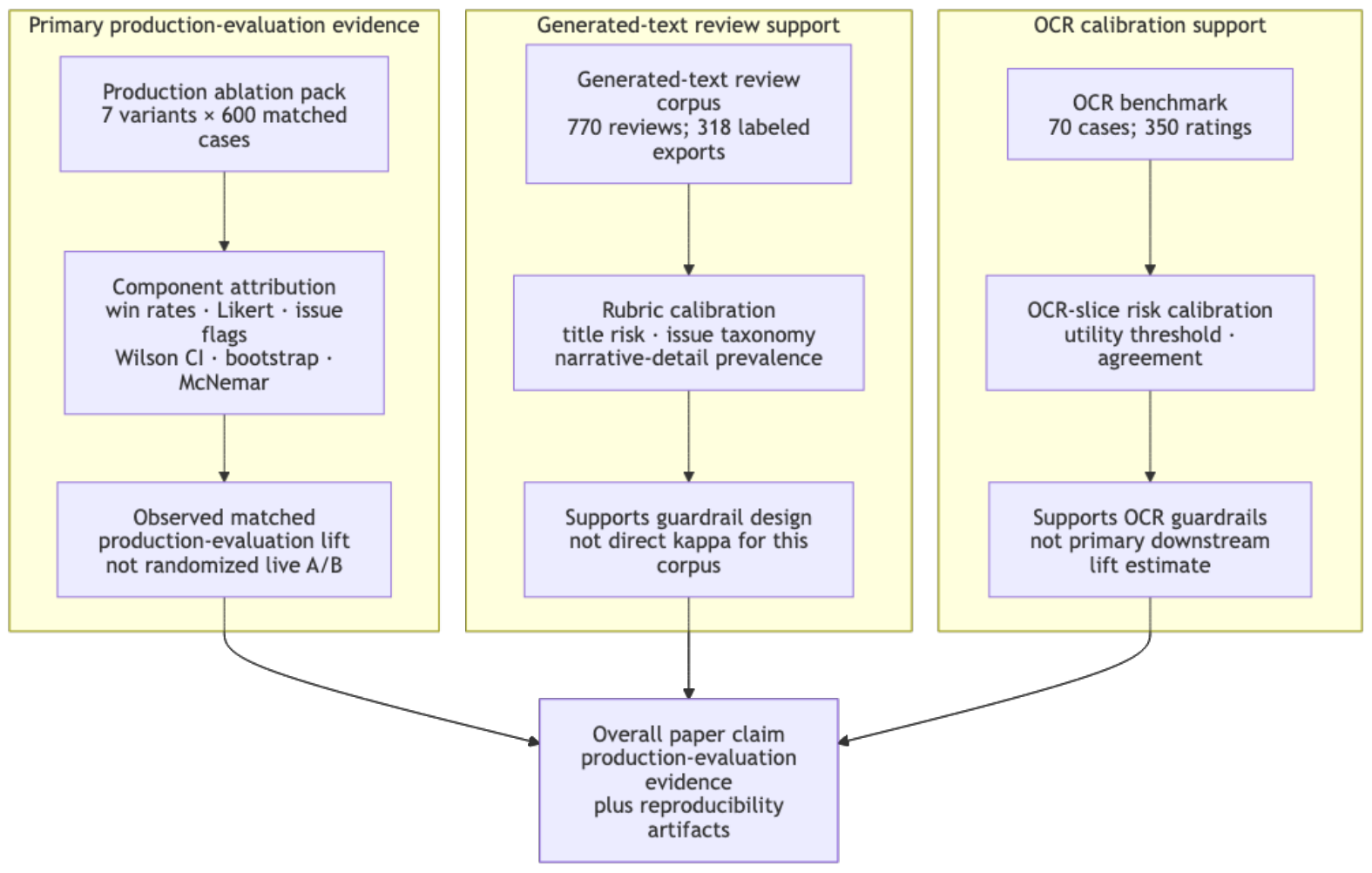}
\caption{Evidence stack and claim boundary. The production ablation is the primary source for lift claims, while the text-review and OCR artifacts calibrate rubrics, guardrails, and risk interpretation.}
\label{fig:evidence-stack}
\end{figure*}

Table~\ref{tab:sources} lists the evaluation evidence. The ablation pack is the primary production-evaluation evidence: it provides matched variants over the same cases, enabling component attribution across a static prompt, manual prompt iteration, bandit-only, mutation-only, single-loop, and full dual-loop configurations. The text and OCR review artifacts are used as supporting calibration evidence for rubrics, guardrails, title-risk interpretation, reviewer drift, and OCR-slice risk. Figure~\ref{fig:evidence-stack} makes this claim boundary explicit.

\begin{table*}[t]
\caption{Evaluation evidence and how it is used. The cleaned evidence package separates export provenance from evaluation-basis labels and uses anonymized proprietary-safe fields.}
\label{tab:sources}
\small
\begin{tabularx}{\textwidth}{p{0.21\textwidth}p{0.23\textwidth}p{0.21\textwidth}X}
\toprule
Source & Sample & Key fields & Use in paper \\
\midrule
Generated-text review artifact & 770 unique reviews; 318 labeled exports; 5 reviewers & 1--5 rating, positive feedback, dispute reason, 22 issue flags, reviewer details, title, summary, evidence excerpt & Human-review calibration, issue taxonomy, title-template risk, dispute mix, narrative-detail prevalence. \\
OCR review artifact & 70 cases; 5 analysts; 350 ratings & Text accuracy, word/line integrity, noise vs. non-text confusion, overall OCR, evidence utility, pass/fail & OCR slice calibration, pass/fail rule, pairwise weighted \(\kappa\), Fleiss' \(\kappa\), OCR risk categories. \\
Production ablation pack & 4,200 rows: 7 variants \(\times\) 600 matched cases; 1,000 amount bootstraps per variant; 1,000 paired bootstraps per comparison & Downstream win, amount-weighted win, Likert score, guardrail pass, issue flags, prompt family/version, mutation count & Component ablations, confidence intervals, pairwise tests, reproducibility schema, and production lift estimates. \\
\bottomrule
\end{tabularx}
\end{table*}

\subsection{Control setup and baselines}
The production ablation uses 600 matched cases shared across all variants. Each case preserves the same hashed case identifier and disputed amount across variants, so pairwise amount-difference bootstraps compare like with like. The static control disables routing, mutation, human feedback, and automated judging. The manual iteration baseline uses human feedback without online bandit routing or automated judging. The bandit-only baseline enables online family/version routing with automated-judge reward but no prompt mutation or human loop. The mutation-only baseline enables mutation with human feedback but no bandit or automated judge. Two single-loop variants enable the online router and mutator with exactly one feedback channel: human-only or auto-judge-only. The full \hopm variant enables bandit routing, mutation, human feedback, and automated judging.

\subsection{Metrics and tests}
We report four outcome families. \emph{Count win rate} is the fraction of cases with a downstream win. Each count rate uses a 95\% Wilson interval~\cite{wilson1927probable}. Because variants are matched at case level, pairwise count tests use exact McNemar tests over discordant cases, with paired bootstrap intervals for count lift. \emph{Amount-weighted win rate} is total disputed amount won divided by total disputed amount; intervals use a nonparametric case-level bootstrap with 1,000 draws~\cite{eves1993bootstrap}. \emph{Reviewer quality} uses mean 1--5 Likert score, positive feedback rate, issue-flag rate, and guardrail-pass rate. Agreement metrics use weighted Cohen's \(\kappa\) for ordinal pairwise OCR ratings~\cite{cohen1968weighted}, Cohen's \(\kappa\) for pass/fail agreement~\cite{cohen1960coefficient}, and Fleiss' \(\kappa\) for multi-rater OCR-slice agreement~\cite{fleiss1971measuring}.

\section{Supporting Review Calibration}
\subsection{Text-review calibration}
The generated-text review deck covers 770 unique reviews, with 318 labeled exports available for estimation. Projected to the 770-review corpus, ratings are concentrated at 3/5: 574 reviews are rated 3, 133 are rated 4, 24 are rated 5, 27 are rated 2, and 12 are rated 1. Thus, 157 reviews (20.4\%) fall in the 4--5 range and 39 reviews (5.1\%) fall in the 1--2 range. Reviewer severity is modest but material: anonymized reviewer means span 2.78--3.37 against a corpus mean of 3.17.

Structured issues are sparse but concentrated. The most frequent projected issue flags are title-not-aligned (29 reviews; 3.8\%) and title-inaccurate (27 reviews; 3.5\%). Summary inaccurate, buyer-ID inaccurate, evidence-not-aligned, inappropriate language, seller-as-buyer, and summary grammar each remain below 1\% at corpus scale. Narrative reviewer detail appears in about 574 reviews (74.5\%), compared with 77 reviews (10.1\%) with any structured issue flag, motivating the dual-loop design.

The corpus is dominated by significantly-not-as-described (SNAD) disputes: approximately 753 of 770 reviews are SNAD, 12 are fraud, and 5 are item-not-received. This concentration justifies specialized title/reason guardrails while also limiting claims about rare dispute categories.

\subsection{OCR calibration}
The OCR benchmark contains 70 cases scored by five analysts, producing 350 ratings. The pass/fail rule is explicit: a case passes when overall OCR is at least 4 and evidence utility is at least 3. Pairwise agreement is moderate-to-strong for ordinal Overall OCR scores, ranging from weighted \(\kappa=0.646\) to 0.753, and stronger for pass/fail agreement, ranging from \(\kappa=0.72\) to 0.85. Fleiss' \(\kappa\) declines with noisier OCR slices: 0.68 for clean chat screenshots, 0.65 for email/page captures, 0.63 for mixed-UI screenshots, 0.59 for label/receipt photos, and 0.54 for handwritten/blurry captures. Pass rates follow the same risk order: 95\%, 85\%, 78\%, 60\%, and 45\%, respectively.

\section{Observed Production-Evaluation Ablation Results}
\subsection{Variant-level outcomes}
Table~\ref{tab:ablations} reports the matched production ablation. The full dual-loop configuration has the highest count win rate (45.7\%, 95\% Wilson CI [41.7, 49.7]\%) and amount-weighted win rate (41.4\%, 95\% bootstrap CI [31.8, 50.8]\%). The static control is lowest on both primary endpoints: 34.7\% count win rate and 22.3\% amount-weighted win rate. The mean Likert score increases from 3.18 under static control to 4.40 under full dual-loop \hopm, while the issue-flag rate falls from 15.3\% to 5.2\%.

\begin{table*}[t]
\caption{Observed production-evaluation ablation results. CI = 95\% interval; count intervals use Wilson intervals and amount intervals use 1,000 nonparametric bootstrap draws.}
\label{tab:ablations}
\scriptsize
\begin{tabularx}{\textwidth}{p{0.18\textwidth}r p{0.13\textwidth} p{0.13\textwidth}rrrrX}
\toprule
Variant & \(n\) & Count win [CI] & Amount win [CI] & Likert & Positive & Issue & Guardrail & Enabled modules \\
\midrule
Static control & 600 & 34.7\% [31.0, 38.6] & 22.3\% [16.7, 29.0] & 3.18 & 32.3\% & 15.3\% & 68.7\% & none \\
Manual prompt iteration & 600 & 37.2\% [33.4, 41.1] & 27.8\% [20.8, 36.1] & 3.42 & 48.8\% & 14.2\% & 70.7\% & human feedback only; no bandit, no automated judge \\
Bandit-only & 600 & 39.2\% [35.3, 43.1] & 30.7\% [23.4, 38.7] & 3.54 & 54.5\% & 9.3\% & 75.7\% & bandit + automated judge; no mutation, no human feedback \\
Mutation-only & 600 & 40.5\% [36.6, 44.5] & 33.2\% [24.9, 41.6] & 3.68 & 64.8\% & 6.8\% & 79.7\% & mutation + human feedback; no bandit, no automated judge \\
Human-only single loop & 600 & 42.2\% [38.3, 46.2] & 35.9\% [27.2, 46.1] & 3.97 & 81.0\% & 5.5\% & 79.0\% & bandit + mutation + human feedback \\
Auto-judge-only single loop & 600 & 41.3\% [37.5, 45.3] & 34.7\% [26.4, 44.5] & 3.82 & 73.2\% & 9.2\% & 76.5\% & bandit + mutation + automated judge \\
Full dual-loop \hopm & 600 & 45.7\% [41.7, 49.7] & 41.4\% [31.8, 50.8] & 4.40 & 96.7\% & 5.2\% & 78.8\% & bandit + mutation + human + automated judge \\
\bottomrule
\end{tabularx}
\end{table*}

\subsection{Pairwise lift and statistical tests}
Table~\ref{tab:pairwise} reports pairwise lifts using the matched case structure. Full dual-loop \hopm outperforms static control by +11.0\,pp in count win rate (paired 95\% bootstrap CI [8.0, 14.2] pp; McNemar \(p=1.31\times10^{-11}\)) and +19.1\,pp in amount-weighted win rate (95\% paired bootstrap CI [10.3, 28.9] pp). Full dual-loop \hopm also improves count lift over both single-loop variants: +3.5\,pp over human-only (McNemar \(p=0.0257\)) and +4.3\,pp over auto-judge-only (McNemar \(p=0.00156\)). Amount-weighted lift over auto-judge-only is positive (+6.7\,pp; 95\% CI [1.1, 14.1] pp), suggesting that human feedback adds value even when the automated judge is active.

\begin{table}[t]
\caption{Selected paired ablation tests. Count lift uses case-paired bootstrap intervals and exact McNemar tests over discordant cases; amount lift uses paired bootstrap intervals.}
\label{tab:pairwise}
\small
\begin{tabularx}{\columnwidth}{Xrrr}
\toprule
Comparison & Count lift [CI] & McNemar \(p\) & Amount lift [CI] \\
\midrule
Manual vs. static & +2.5 [0.0, 5.0] & 0.0769 & +5.5 [-1.2, 13.4] \\
Bandit-only vs. static & +4.5 [1.5, 7.5] & 0.00451 & +8.4 [1.6, 16.8] \\
Mutation-only vs. static & +5.8 [2.7, 9.0] & 0.000366 & +10.9 [2.1, 21.3] \\
Human-only vs. static & +7.5 [4.2, 11.0] & \(1.60\times10^{-5}\) & +13.6 [4.1, 24.4] \\
Auto-only vs. static & +6.7 [3.5, 9.8] & \(5.46\times10^{-5}\) & +12.4 [3.0, 22.0] \\
Full \hopm vs. static & +11.0 [8.0, 14.2] & \(1.31\times10^{-11}\) & +19.1 [10.3, 28.9] \\
Full \hopm vs. human-only & +3.5 [0.5, 6.5] & 0.0257 & +5.5 [-3.0, 14.2] \\
Full \hopm vs. auto-only & +4.3 [1.8, 7.0] & 0.00156 & +6.7 [1.1, 14.1] \\
\bottomrule
\end{tabularx}
\end{table}

\section{Reproducibility Without Proprietary Data}
The reproducibility package uses structure rather than raw content. It contains four artifacts. First, a schema defines case-level records with stable hashed IDs, variant IDs, prompt family/version, OCR slice, amount bucket, flags, and outcomes. Second, the prompt-token taxonomy in Table~\ref{tab:tokens} defines the mutation surface. Third, the Likert rubric in Table~\ref{tab:rubric} defines reviewer judgment. Fourth, the guardrail taxonomy in Table~\ref{tab:guardrails} maps observed errors to executable checks.

\begin{table}[t]
\caption{Proprietary-safe case schema for reproducing the evaluation design.}
\label{tab:schema}
\small
\begin{tabularx}{\columnwidth}{p{0.28\columnwidth}X}
\toprule
Group & Fields released or reconstructable \\
\midrule
Case identity & \texttt{case\_id\_hash}, \texttt{case\_index}, \texttt{dispute\_reason}, \texttt{amount\_bucket}, \texttt{disputed\_amount\_usd}. \\
Evidence inputs & Normalized message spans, OCR slice, OCR expected quality, evidence-span IDs, masked actor roles, dates, and amounts. \\
Prompt state & \texttt{variant\_id}, prompt family, prompt version, bandit/mutation/human/auto-judge flags, selected-by-bandit flag, mutation-token count. \\
Feedback & Reviewer or judge source, Likert score, positive feedback, narrative-detail flag, structured issue flags. \\
Outcomes & Guardrail pass, downstream win, amount-weighted win numerator, and audit timestamp. \\
\bottomrule
\end{tabularx}
\end{table}

\paragraph{Constructed anonymized example.}
A reproducible public case can be represented as follows without exposing proprietary evidence. The input contains a SNAD dispute, two masked messages (Buyer: ``item powers on but does not synchronize''; Seller: ``please return it for inspection''), a clean-chat OCR slice with expected quality 4.7, and a \$50--\$99 amount bucket. A title-risk-focused prompt version generates: ``Buyer states the item failed to synchronize and seller offered return inspection.'' Guardrails check that the title mentions the SNAD-specific defect rather than generic receipt, roles are preserved, each claim links to a message span, and no unsupported legal intent is asserted. The resulting record contains only hashed case IDs, masked spans, structured flags, and outcome fields.

\begin{table}[t]
\caption{Reviewer Likert rubric used for generated evidence quality.}
\label{tab:rubric}
\small
\begin{tabularx}{\columnwidth}{p{0.11\columnwidth}X}
\toprule
Score & Interpretation \\
\midrule
1 & Unusable: materially wrong, unsafe, or unsupported by supplied evidence. \\
2 & Poor: contains a major title, role, evidence, or summary defect requiring rewrite. \\
3 & Usable with caution: mostly grounded but incomplete, generic, or missing important nuance. \\
4 & Strong: grounded and useful, with only minor wording or completeness issues. \\
5 & Audit-ready: directly supports the dispute decision, preserves roles and remedy, and needs no substantive edit. \\
\bottomrule
\end{tabularx}
\end{table}

\begin{table}[t]
\caption{Guardrail taxonomy derived from observed review flags and OCR evaluation.}
\label{tab:guardrails}
\small
\begin{tabularx}{\columnwidth}{p{0.30\columnwidth}X}
\toprule
Guardrail & Failure detected \\
\midrule
Title not aligned & Title emphasizes a generic event instead of the dispute reason or remedy. \\
Title inaccurate & Title states a fact not supported by evidence. \\
Summary inaccurate & Summary changes, omits, or overstates material evidence. \\
Buyer ID inaccurate & Identifier or actor metadata is copied incorrectly. \\
Evidence not aligned & The cited evidence span does not support the generated claim. \\
Inappropriate language & Output includes unsafe, accusatory, or policy-inappropriate phrasing. \\
Seller as buyer & Buyer/seller roles are reversed. \\
Summary grammar & Output is hard to read, malformed, or grammatically defective. \\
OCR risk & OCR slice or quality score is below the evidence-utility threshold. \\
\bottomrule
\end{tabularx}
\end{table}

\section{Limitations}
The production ablation is a matched production-evaluation analysis, not a randomized traffic A/B test. It substantially strengthens component attribution because all variants are evaluated on the same 600 cases, but it cannot by itself prove live deployment lift under future traffic shifts. The observed text-review artifacts do not preserve stable shared case IDs across reviewers, so direct inter-rater kappa for the text-generation review corpus is not reproducible from those artifacts alone. The dispute mix is heavily skewed toward SNAD, so generalization to fraud and item-not-received disputes should be tested separately. Finally, proprietary data constraints prevent raw evidence release; this is mitigated by a public schema, pseudocode, rubrics, aggregate statistics, and a constructed example, but not eliminated.

\section{Conclusion}
\hopm addresses a practical problem in guardrailed LLM document generation: adapting prompts online while preserving auditability and evidence grounding. The strengthened evaluation reports observed production-evaluation lift, exact baseline definitions, matched 600-case sample sizes, confidence intervals, paired tests, and reproducibility artifacts. In the production ablation, the full dual-loop configuration is the strongest variant by count win rate, amount-weighted win rate, Likert quality, and issue reduction. The next validation step is a live randomized holdout or staggered rollout that compares the full dual-loop policy against a frozen static or single-loop control under the same reviewer and case distribution.

\appendix
\section{Complete Metric Ledger for Public Preprint}
This appendix preserves the non-proprietary metrics used to calibrate the production evaluation while keeping raw evidence, reviewer identities, and customer data out of the public artifact.

\subsection{Generated-text corpus metrics}
The generated-text review artifact contains 770 unique reviews and 318 labeled exports with full fields. The projected corpus mix is approximately 726 core reviews and 44 pilot spot-checks. Narrative detail appears in 574 of 770 projected reviews (74.5\%), while any structured issue flag appears in 77 of 770 projected reviews (10.1\%). The projected dispute mix is 753 SNAD reviews, 12 fraud reviews, and 5 item-not-received reviews. The main title-risk pattern is generic receipt emphasis: one recurring template projects to 85 reviews and roughly 22 title-issue cases, while a smaller return-instruction template projects to 15 reviews.

\begin{table}[h]
\caption{Projected generated-text corpus distribution.}
\label{tab:appendix_text}
\small
\begin{tabularx}{\columnwidth}{Xr}
\toprule
Metric & Value \\
\midrule
Unique reviews represented & 770 \\
Labeled reviewer exports & 318 \\
Reviewers & 5 \\
Projected rating 1 / 2 / 3 / 4 / 5 & 12 / 27 / 574 / 133 / 24 \\
Projected 4--5 ratings & 157 (20.4\%) \\
Projected 1--2 ratings & 39 (5.1\%) \\
Projected narrative-detail reviews & 574 (74.5\%) \\
Projected issue-flagged reviews & 77 (10.1\%) \\
Title-not-aligned projected count & 29 (3.8\%) \\
Title-inaccurate projected count & 27 (3.5\%) \\
Projected SNAD / fraud / INR mix & 753 / 12 / 5 \\
Projected corpus mean & 3.17 \\
Anonymized reviewer mean range & 2.78--3.37 \\
Core projected positives / issues & 593 / 63 \\
Pilot projected positives / issues & 22 / 15 \\
\bottomrule
\end{tabularx}
\end{table}

\subsection{Small-overlap human calibration slice}
A separate human-evaluation review slice contains 10 cases, 7 analysts, and 61 ratings. It is useful for calibration but is not the primary production-lift evidence. Global multi-rater agreement is near zero in this tiny slice (Overall Fleiss' $\kappa=0.030$; pass/fail Fleiss' $\kappa=-0.007$), which is consistent with small overlaps and imbalanced labels. The global weighted median of per-case Overall means is 3.571. Anonymized top pairwise weighted-$\kappa$ alignments are 0.727 ($n=10$), 0.647 ($n=9$), and 0.571 ($n=9$). Two case fact-sheet examples have Overall means 4.33 $\pm$ 0.52 with pass rate 1.00 and 3.00 $\pm$ 1.26 with pass rate 0.50.

\subsection{OCR benchmark metrics}
The OCR review artifact contains 70 cases, 5 analysts, and 350 ratings across clean chat, email/page, mixed-UI, label/receipt, and handwritten/blurry slices. The global weighted median per-case Overall OCR score is 3.2. Pairwise weighted-$\kappa$ for Overall OCR ranges from 0.646 to 0.753 in the supplied benchmark, with top anonymized alignments 0.753, 0.741, and 0.722. Reviewer disagreement concentrates in blur drop-out (mean case SD 0.91), cropped text loss (0.90), and UI bleed/initial capture (0.78). A passing clean-chat example scores 4.40 $\pm$ 0.55 with pass rate 1.00; a failing analyzer-error label/receipt example scores 1.00 $\pm$ 0.00 with pass rate 0.00.

\subsection{Production claim boundary}
The primary causal-strength claim is component attribution under a matched production-evaluation design. Each ablation variant is evaluated on the same 600 cases, and pairwise tests exploit this matching. The design supports the statement that full dual-loop \hopm performs best among the evaluated configurations on the matched production-evaluation set. It does not by itself establish future randomized traffic lift under all traffic distributions; that requires a live holdout or staggered rollout.

\bibliographystyle{ACM-Reference-Format}
\bibliography{refs}
\end{document}